\pdfoutput=1
\documentclass[sigconf]{acmart}

\usepackage{booktabs} 
\usepackage{enumitem}
\usepackage{amsmath}
\usepackage{epstopdf}
\usepackage{xcolor,colortbl}
\usepackage{algorithm}
\usepackage[noend]{algpseudocode}
\graphicspath{{figures/}}

\author{Samira Briongos}
\affiliation{
  \institution{Universidad Polit{\'e}cnica de Madrid}}
\email{samirabriongos@die.upm.es}

\author{Gorka Irazoqui}
\affiliation{
  \institution{Worcester Polytechnic Institute}}
\email{girazoki@wpi.edu}

\author{Pedro Malag{\'o}n}
\affiliation{
  \institution{Universidad Polit{\'e}cnica de Madrid}}
\email{malagon@die.upm.es}

\author{Thomas Eisenbarth}
\affiliation{
  \institution{Worcester Polytechnic Institute}}
\email{teisenbarth@wpi.edu}

\setcopyright{none} 
\renewcommand\footnotetextcopyrightpermission[1]{} 
\newcommand{\CS}{CacheShield}
\begin{document}
\title{CacheShield: Protecting Legacy Processes Against Cache Attacks}

\begin{abstract}
Cache attacks pose a threat to any code 
whose execution flow or memory accesses depend on sensitive information. 
Especially in public clouds, where caches are shared across several tenants, cache attacks remain an unsolved problem.
Cache attacks rely on evictions by the spy process, which alter the execution behavior of the victim process. 
We show that hardware performance events of cryptographic routines reveal the presence of cache attacks. Based on this observation, we propose \CS , a tool to protect legacy code by monitoring its execution and detecting the presence of cache attacks, thus providing the opportunity to take preventative measures. \CS\ can be run by users and does not require alteration of the OS or hypervisor, while 
previously proposed software-based countermeasures require cooperation from the hypervisor. 
 Unlike methods that try to detect malicious processes, our approach is lean, as only a fraction of the system needs to be monitored. It also integrates well into today's cloud infrastructure, as concerned users can opt to use \CS\ without support from the cloud service provider.
Our results show that \CS\ detects cache attacks fast, with high reliability, and with few false positives, even in the presence of strong noise.
\end{abstract}


\maketitle

\section{Introduction}

Modern computing technologies like cloud computing build on shared hardware resources opaquely accessible by independent tenants, ensuring protection through sandboxing techniques.  However, although this isolation is solid at the logical level, ensuring tenants cannot access each others memory, hypervisors cannot properly prevent information leakage stemming from the shared hardware resources such as caches. Compared to other resources like Branch Prediction Units or the DRAM, caches can be exploited to recover fine grain information from co-resident tenants in shared environments. 

Cache attacks extract private information by setting up the cache memory,  executing the victim process and observing effects related to sensitive data. 
Time-based cache attacks measure the effect on the victim process execution time \cite{Bernstein05}
while access-based cache attacks measure the effect on the attacker \cite{Percival05cachemissing}.
Practical access-based cache attacks have been published for cloud environments with different variants: \texttt{Prime+Probe}
\cite{Ristenpart09,DBLP:conf/ccs/ZhangJRR12,ZhangEtAlPaaS,sca,lastlevel}, 
\texttt{Flush+Reload}
\cite{GullaschEtAl2011,YaromEtAl2014,Irazoqui14,Gulmezoglu15}, and  \texttt{Flush+Flush}
\cite{GrussEtAl2016}. All of them have demonstrated to recover cryptographic keys, break security protocols or infer privacy related information. They have shown that attacks can succeed in contemporary public cloud systems, with severe consequences to sensitive data of cloud customers.

To deter cache attacks, several techniques for detection and/or mitigation have been proposed.
Most of the proposed mitigation techniques succeed in stopping cache based attacks, but are not being adopted by cloud service providers. Proposed hardware countermeasures require making modifications to the hardware that not only induce severe performance penalties but also take years to integrate and deploy into the infrastructure. Cloud hypervisors, on the contrary, can implement any of the proposed hypervisor based countermeasures~\cite{180212,li2014stopwatch,Varadarajan14} by just making small modifications to the kernel configuration. Despite the immediate fix that these countermeasures would provide, they are not being adopted by cloud providers, mainly due to the constant performance overhead that they add to their systems. Other feasible mitigation proposals consider periodic VM migration to avoid long-term co-location. VM migration, however, also  introduces extra overhead whether there is
an attack or not. Other proposals suggest that just as the attacker uses a side-channel to obtain information,
the VM can defend itself by using a side-channel to detect co-resident tenants with possibly malicious intentions \cite{HomeAlone11}

The current situation leaves tenants with little help from hardware and hypervisor designers or cloud service providers to protect themselves against cache attacks. Thus, we observe the necessity of giving those tenants that \emph{voluntarily} want to protect against cache attacks, tools to defend themselves. So far, all known cache attacks have in common that they cause
cache misses \emph{in the victim VM process}. Thus, detecting an anomaly in the number of cache misses in the victim can indicate an ongoing cache attack and thus trigger
VM migration or other actions to mitigate the attack.

Cache misses can be obtained by reading the hardware performance counters found in all modern processors. These hardware event counters track hardware events such as cache misses, and were originally intended to enable the detection of bottlenecks in executed software. Optimization is not the only application of these counters, it has also been demonstrated that the hardware performance counters are also useful to detect malware and security breaches \cite{6560672,DemmeEtAl2013,BahadorEtAl2014,TangEtAl2014}. Libraries such as PAPI (Performance Application Programming Interface), facilitate the task of configuring and reading those hardware counters.

There have been several attempts to detect cache attacks using the
hardware counters~\cite{ChiappettaEtAl2016,Payer2016,Zhang2016}, but they have strong drawbacks.
Some works require the hypervisor to periodically monitor all existing
processes, which introduces a great overhead in CPU usage and depends on how efficiently an attacker can hide from the monitoring tool~\cite{Zhang2016,Payer2016}.
Other works offer solutions applicable to multi-process environments,
but not feasible in cloud environments~\cite{ChiappettaEtAl2016,GrussEtAl2016}. 

We propose to use a monitoring service inside the VM that detects anomalies
in the cache miss hardware performance counter \emph{only in the victim side}.
The monitoring service can be activated on demand inside the VM.
The performance counters must be exposed to the VM in order to be feasible. Just changing the configuration of the hypervisor, it is possible to enable performance counters access inside the VM. This access can be enabled only in the VMs that request the service, and as the hypervisor is responsible for the virtualization of the counters that can be read inside a VM, they refer uniquely to this VM. That is, one VM can not read counters referring to another VM, even when they share the hardware. Right now, most cloud service providers only expose them if the customer is renting the whole machine, probably due to their fear of utilization as a side channel in hardware shared by various tenants. However, we believe they do not have much to worry about, as current attacking techniques exploiting shared hardware expose much more information than hardware counters would. 

Our work demonstrates for the first time, that performance counter access for tenant VMs can indeed be utilized to \emph{improve security} of the tenants. We offer tenant VMs a new monitoring service, \CS,  to detect cache attacks. \CS\ can be activated before running sensitive processes. \CS\ detects attacks quickly and with high reliability and low CPU overhead, due to the use of Page's cumulative sum method~\cite{page1954continuous}. The CUSUM method is an unsupervised anomaly detection method, ensuring that even new attack techniques can be detected with high confidence.
\CS\ automatically turns off if the monitored process is idle by detecting the lack of
activity, resulting in a significant reduction in CPU processing overheads.
In summary, our work

\begin{itemize} [leftmargin=5ex]
\item presents a performance counter based monitoring service that users can voluntarily activate to detect when they are under attack.
\item only monitors the victim process upon when active, i.e., the cloud service provider does not waste cycles continuously monitoring all processes.
\item only requires the hypervisor to enable VM access to the performance counters, a feature commonly supported by all major hypervisor systems, including KVM, VMware and Xen. No other additional help from the underlying system is needed.
\item implements an efficient algorithm that maximises fast and reliable attack detection, while minimizing false positives and keeping the performance overhead minimal and restricted to the victim VM.
\item succeeds detecting all existing cache attacks, including stealthy attacks that are miss-detected by other solutions, e.g. \texttt{Flush+Flush}, since our detection uses attack characteristics that are independent of attack and victim behavior.
\end{itemize}

The rest of the paper is organized as follows. After discussing background and related work in Section~\ref{sec:background}, we show that monitoring performance events of a victim process is sufficient for reliable attack detection in Section~\ref{sec:self}. \CS\ is developed in Section~\ref{sec:goals}. Section~\ref{sec:eval} presents the performance evaluation in several relevant scenarios and in section~\ref{sec:counter} we suggest different countermeasures. Finally, Section~\ref{sec:conclusion} discusses the conclusions of our work.	

\section{background and related work}\label{sec:background}
\subsection{Cache attacks}
In the last years cache attacks have shown to pose a big threat in those systems in which the underlying hardware architecture is shared with a potential attacker. Cache attacks monitor the utilization of the cache to retrieve information about a co-resident victim. Indeed, if the utilization of such a hardware piece is directly correlated with a security-critical piece of information (e.g., a cryptographic key) the consequences of the attack can be as devastating as an impersonation of the victim.

Two main cache attack designs out-stand over the rest: the \texttt{Flush+Reload} and the \texttt{Prime+Probe} attacks. The first was first introduced in~\cite{GullaschEtAl2011}, and was later extended to target the LLC to retrieve cryptographic keys, TLS protocol session messages or keyboard keystrokes across VMs~\cite{YaromEtAl2014,GrussEtAl2015,IrazoquiEtAl2015}. Further, Zhang et al.~\cite{ZhangEtAlPaaS} showed that \texttt{Flush+Reload} is applicable in several commercial PaaS clouds. Despite its popularity and resistance to micro-architectural noise, the \texttt{Flush+Reload} presents a main drawback, as it can only be applied in systems in which memory deduplication mechanisms are in place, and further, can only recover information coming from statically allocated data.

The \texttt{Prime+Probe} attack design, contrary to the \texttt{Flush+Reload} attack, is agnostic to special OS features in the system, and therefore it can not only be applied in virtually every system, but further, it can additionally recover information from dynamically allocated data. This attack was first proposed for the L1 data cache in~\cite{OsvikEtAl2006}, while later was expanded to the L1 instruction cache~\cite{aciiccmez2008}. Recently, it has been shown to also bypass several difficulties to target the LLC and recover cryptographic keys or keyboard typed keystrokes~\cite{lastlevel,sca,DBLP:conf/uss/LippGSMM16}. Even further, the \texttt{Prime+Probe} attack was used to retrieve a RSA key in the Amazon EC2 cloud~\cite{Inci2016}.

Variations of both attacks have also been proposed to bypass specific difficulties found in some systems (e.g., lack of a flush instruction in the Instruction Set Architecture). Perhaps the one that most directly influences this work is the design of the \texttt{Flush+Flush} attack, as it was proposed to be stealthy and bypass attack monitoring systems~\cite{GrussEtAl2016}. This attack retrieves information by measuring the execution time of the flush instruction, thus avoiding direct cache accesses. As we will see, although this design might be effective against some of the proposed detection systems, ours correctly identifies when such an attack is being executed.

\subsection{Performance counters}

The performance counters are special purpose hardware registers that count a broad spectrum of low-level hardware events related to code execution. The selection of observable events is usually larger than the number of actual counters, hence, counters must be configured in advance. All events associated with a counter are recorded in parallel. As the PMU allows detailed insight into the state of the processor in real-time, it is a valuable tool for debugging applications and their performance. The list of available events consequently focuses on waiting periods (e.g. clock cycles the processor is stalled), memory or bus accesses (e.g. cache misses or DRAM requests), and other performance-critical metrics like branch prediction or TLB events.

All main micro-processor architectures, i.e., Intel, AMD and ARM, include a bigger or a smaller number of these configurable registers. However, while monitoring of these hardware events in Intel and AMD processors is usually possible from user mode (when referring to an application also being run in user mode), ARM devices require root rights to enable them. Emulating the behavior in ARM devices, cloud providers might disable the utilization of performance counters from guest VMs. Indeed we find two main reasons why they would do this

\begin{itemize}[leftmargin=5ex]
\item Performance counters might be utilized with malicious purposes, similarly to the way the thermal sensor was used in~\cite{190938}, and retrieve information from co-resident user hardware utilization~\cite{Bhattacharya2015}, which in theory should not be possible as the hypervisor only gives information to each VM about itself.
\item As performance counters are hardware dependent, giving a guest VM access to benign utilization of performance counters might be problematic if guest VMs are migrated over different architectures, as customers would have to design code for different hardware architectures.
\end{itemize}

We do not believe that these facts should make cloud providers disable the usage of performance counters from guest VMs, specially when one can use them as a protection mechanism as we will see later in this work. In fact, attackers have already found alternative ways to retrieve the same information performance counters give. For instance, attackers can read the cycle counter or an incremental thread to know when TLB or cache misses occur. Thus, disabling the counters does not entirely prevent the leakage of hardware events information. As for the second claim, a possible solution could be to create clusters with the same hardware configuration, and migrate VMs within this cluster. Thus, we do not believe the above concerns are strong enough arguments against the guest VM performance counter usage. In this paper we will further show that such a usage can indeed offer more protection to cloud infrastructure customers.

\subsection{Detection, mitigation and other countermeasures}

HPCs have been used to detect generic malware~\cite{TangEtAl2014,BahadorEtAl2014,SinghEtAl2017} as well as microarchitectural attacks~\cite{ChiappettaEtAl2016,Zhang2016,GrussEtAl2016,Payer2016}. Their success mostly depends on the ability to correctly identify cache (and other resource) attack patterns monitoring the associated event in the HPC. This approach is usually implemented at the OS or hypervisor level that has enough permissions to monitor what is running in the system. However, we observe two main problems with these detection-based approaches:
\begin{itemize}[leftmargin=5ex]
\item Most of these detection approaches incur severe performance overheads that hypervisors or OSs do not seem willing to pay, as to the best of our knowledge no OS is implementing such a mechanism. This leaves the user of the system with few resources to know whether her code will be executed in a safe environment.
\item As these detection countermeasures base their success on the monitoring of both the victim and the attacker processes, the attacker can vary patterns in a smart way to try to bypass the detection mechanisms.
\end{itemize} 

These facts are observed, for instance in~\cite{Zhang2016,Payer2016,ChiappettaEtAl2016}. All three works incur significant overheads on all applications. CloudRadar, for example, requires three dedicated cores for its detection~\cite{Zhang2016}. In addition, they usually assume the ability to monitor the attacking process~\cite{GrussEtAl2016,ChiappettaEtAl2016}, which is not possible across VM boundaries (except for the hypervisor), and usually not even possible for user-level processes. 

Detection-based countermeasures are not the only possibility shown to prevent cache timing attacks. Preemptive approaches can be taken at the hardware, software and application level. The first usually requires changes in the hardware pieces such that collisions in the cache can not happen, or if they do, they do not carry information~\cite{DBLP:conf/isca/WangL07}. The second involves the utilization of specific software features (e.g., page allocation) to prevent two processes from colliding in the cache~\cite{180212}. Finally, the latter is achieved by utilizing specific tools to ensure a security sensitive binary does not leak information, even if it is under attack~\cite{ZanklEtAl2017}.

\section{Victim-based Attack Detection}\label{sec:self}

Our objective is to build an attack detection tool that detects any abuse of the LLC without any modifications to the hypervisor, OS, or the CPU hardware. Unlike previous approaches, we show that monitoring the behavior of a victim application is sufficient for the detection of cache attacks. To that end, we first analyze the behavior of these \emph{victim applications} by monitoring several critical hardware performance counters. This behavior of critical applications is analyzed in the presence and absence of various cache attacks, and further analysis is performed to determine how well each counter serves as an indicator for ongoing attacks. 

For the sake of simplicity, we base our analysis on cryptographic algorithms, which are the most popular target for cache attacks. Our approach can also detect attacks on other security-critical pieces of code like SSL/TLS protocol stacks. There are different types of cryptographic algorithms in use, which traditionally have been classified as symmetric cryptography an public-key cryptography.

\begin{description}[leftmargin=4ex]
\item[Symmetric cryptosystems] are sometimes also called private key algorithms, and include algorithms for encryption, authentication as well as hashing. Encryption and authentication schemes use single key for both the encryption/authentication and decryption/verification. Popular algorithms include AES and DES for encryption, SHA-2 and SHA-3 for hashing and HMAC or GCM for authentication or authenticated encryption. Symmetric primitives are usually heavily optimized for performance and feature constant execution flows. However, some implementations make use of table look-ups, which often result in exploitable cache leakage. One example is AES, the most widely used encryption algorithm. For AES, table look-ups are difficult to avoid, unless hardware support such as AES-NI is available.

\item[Public key cryptosystems] use a public key for encryption or verification and private key for decryption or signing. While public key cryptography can be used in more flexible ways, the used primitives are much more costly than for symmetric cryptography. As a result, public key cryptography is mainly used for authentication and key exchange to establish a communication session, where payloads are protected using symmetric cryptography. Another important offered service are certificates, which require digital signatures for generation and verification of certificates.  RSA, ECC and ElGamal are currently the prevailing schemes for public key cryptography.
\end{description}

As explained before, we only collect information about the victim processes, i.e. the processes operating on sensitive data. This approach avoids the need of monitoring other processes or VMs running in the same host. This approach also avoids relying on the information gathered from an attacker who might try to hide changes on its behavior to avoid triggering an alarm. 
Considering that each kind of algorithm presents different characteristics, we gather and analyze data of the execution of different algorithms in an initial scenario. Next, we show that the main results obtained in this scenario can be extended to others. 

\subsection{Analyzing Hardware Performance Events}

\definecolor{Gray}{gray}{0.85}
\newcolumntype{a}{>{\columncolor{Gray}}c}

\begin{table*}[tbh]
\caption{Overview of most relevant hardware performance counters in the presence and absence of attacks, over 1 million calls to RSA and AES, as well as their rankings according to the \texttt{InfoGain} and \texttt{relief} metrics. Level 3 cache misses, PAPI\_L3\_TCM, clearly have the strongest information for cache attacks.\label{tab:HPCEval}}
\vspace{-3mm}
  \begin{tabular}{lccacaccacc}
	\toprule
  \textbf{Performance}  &\textbf{AES} &\multicolumn{4}{c}{\textbf{AES w/ Attack}} &\textbf{RSA} & \multicolumn{2}{c}{\textbf{RSA}} & \multicolumn{2}{c}{\textbf{Joint Evaluation}}\\ 
   \textbf{Counter }      & \multicolumn{1}{c}{\textbf{Normal}} & \multicolumn{2}{c}{\textbf{1 line}}& \multicolumn{2}{c}{\textbf{4 lines}} & \multicolumn{1}{c}{\textbf{Normal}} & \multicolumn{2}{c}{\textbf{Attack}} & \multicolumn{2}{c}{\textbf{Algorithms}}\\ 
  & $\mathbf{\mu_{n}}$  & $\mu_{a1}$  & \multicolumn{1}{c}{$\mu_{a1}-\mu_{n}$} & $\mu_{a2}$ & \multicolumn{1}{c}{$\mu_{a2}-\mu_{n}$}  & $\mu_{n}$ & $\mu_{a}$ & \multicolumn{1}{c}{$\mu_{a}-\mu_{n}$} & \texttt{infoGain} & \texttt{Relief} \\
\midrule
  \textbf{PAPI\_L3\_TCM}  & 0.0002 & 0.92 & 0.9189 & 3.56 & 3.5598 & 1.12 & 2601.4 & 2600.28 & 0.885 & 0.245 \\
  \textbf{Cycles (rdtsc)}  & 612.33 & 828.60 & 216.27 & 1151.71 & 539.38 & 8.840e+07 & 8.956e+07 & 1.151e+06& 0.714 & 0.014 \\ 
  \textbf{PAPI\_REF\_CYC}  & 61.93 & 71.01 & 9.08 & 79.93 & 18 & 2.453e+06 & 2.484e+06 & 3.1e+04 & 0.683 & 0.005 \\
  \textbf{PAPI\_CA\_SNP}  & 21.95 &28.77 & 6.82 &30.35 & 8.4 & 727.87 &3417.5 & 2689.63 & 0.531 & 0.034 \\
  \textbf{PAPI\_CA\_INV}  & 21.99 & 28.83 & 6.86 & 30.45 & 8.46 & 727.88 & 3417.6 & 2689.72 & 0.530 & 0.033\\
  \textbf{PAPI\_L3\_TCR}  & 24.65 & 24.73 & 0.08 & 25.90 & 1.25 & 490.47 & 3253.6 & 2763.13 & 0.528 & 0.029 \\
  \textbf{PAPI\_L2\_TCM}  & 28.31 & 28.51 & 0.2 & 28.42 & 0.09 & 559.27 & 3325.6 & 2766.33 & 0.513 & 0.028 \\
  \textbf{PAPI\_L2\_ICM}  & 15.51 & 9.16 & -6.35 & 11.86 & -3.65 & 381.12 & 3149.2 & 2768.08 & 0.510 & 0.056 \\  
  \bottomrule
\end{tabular}
\vspace{-3mm}
\end{table*}

Modern server CPUs make a large number hardware performance counters available, but only a limited number, typically 4 to 8, can be monitored in parallel. We use the Performance API (PAPI)~\cite{Mucci99papia} to access the performance counters. PAPI provides sufficient resolution to detect attacks while it also simplifies the task of collecting performance data. In this preliminary step, we collect data from 30 accessible hardware event counters on our test platform, for sample public and private key algorithms, in the presence and the absence of cache attacks. The PAPI interface provides instructions that allow us to read the counters for our process before and after each cryptographic operation, that is, we get detailed information about the variation of the counters for a single encryption or decryption execution. Since the number of counters that can be read at the same time is limited, we collect the data for different groups of counters at different times. We then join the data and compute the statistics. Once we get all this data, we carry further study to determine and quantify which counters provide meaningful information to detect the attacks.

As sample victim algorithms for this analysis, we chose the software AES T-Table implementation and the RSA sliding window  implementation (with flag RSA\_FLAG\_NO\_CONSTTIME set) of OpenSSL 1.0.1f, which give representative results for public key and symmetric key cryptography. As sample attacks we use 
the \texttt{Flush+Reload} against both implementations. \texttt{Flush+Reload} tries to gain information from the execution of certain instructions or from the accesses to certain data which depend on the key. For the used version of RSA, attacks target the instructions (depending on the implementation RSA can also be attacked considering accesses to data), while AES attacks are an example of cache attacks focused on the data.

Our experiments are performed on an Intel Core i7-4790 CPU 3.60\,GHz machine with 8\,MB of L3 cache and 8\,GB of RAM, with Centos 7 OS. For each counter we collect samples for 1 million encryption or decryption operations. One noteworthy observation is that, whereas in the case of AES the values of the counters do not seem to depend on the key. For the analyzed vulnerable RSA implementation, however, some of the parameters depend on the value of the key.
This behavior can be noticed, for example, in the number of instructions executed and in the decryption times. In fact, the number of operations performed depends on the distribution of zeros and ones in the key. However, while the number of instructions is not affected by the attacks, the decryption times are, as they include the extra times for cache misses. 

We can select up to 5 or 6 counters which are representative of the attacks, as this is the maximum number of counters readable in parallel on our platform. The number of counters that can be read at the same time also varies depending on which counters are used and the combination of them. In order to decide which counters carry more information relative to the attacks, we use the WEKA tool~\cite{Weka_article}. This tool was designed with the aim of allowing researchers to easily access to state-of-the-art techniques in machine learning. WEKA implements several algorithms to perform attribute selection. As inputs for the tool, we select a subset among all the samples (otherwise the time it takes to perform the selection increases exponentially). We randomly select 50000 instances of each of the groups, that is for AES attack and non-attack and for RSA attack and non-attack, so we obtain 200000 samples with information about 30 counters, each labeled with '1' for attacks and '0' for non-attacks.

We first use the \texttt{infoGain} function, which evaluates the worth of an attribute by measuring the information gain with respect to the class according with ``\texttt{InfoGain}(Class,Attribute) = H(Class) - H(Class | Attribute)'', where H is the entropy. Note that our experiments are balanced between attack and non-attack "classes'', that is H(Class)$=1$, thus an ideal attribute would gain 1 bit. Values around 0.5 may indicate the attribute carries meaningful information, but only for one of the algorithms or one of the attacks. Thus, L3 cache misses are not only the most meaningful predictions, but also work across the considered scenarios.

We have also evaluated the relief algorithm~\cite{kira1992feature} for feature selection. Unlike the InfoGain, which only evaluates information gained from each attribute individually, the relief algorithm outputs a score of the predictive value of an attribute relative to other attributes. More positive weights indicate more predictability for this attribute. To calculate the weight of an attribute, it iteratively first identifies the nearest neighbors from the same and different classes. Then, weight increases if a change in the attribute leads to a change in the class and decreases when a change in the attribute value has no effect on the class. 

Table~\ref{tab:HPCEval} presents a summary of the counters which give most relevant information for detection according to the selection algorithms, altogether with their mean values for the considered scenarios, and with the differences between attacks and the \emph{expected} behavior.
Both tests indicate that L3 cache misses are most meaningful. In fact, the relief algorithm scores all other attributes with very low scores, implying only little additional gain from using them.  

\subsection{Concurrent Signal Assessment}

Tracking hardware performance events for each cryptographic operation showed that victim-based attack detection is feasible and helped identifying relevant counters.
However, achieving fast detection with this approach, would require adding instructions in the middle of the code we want to protect.  
Hence, it requires alteration of the target code, which adds unnecessary burden on the user and diminishes practicality. Also, for more effective attack detection, it is preferable to read performance counters concurrently to the execution of the sensitive process. This way, even attacks that succeed during the execution of a single call to the sensitive function, e.g. the attacks presented in~\cite{eprint-2014-25799,YaromEtAl2014}, can be detected and prevented in time. 

\begin{figure}[tb]
\centering
\includegraphics[width=0.49\textwidth,trim={45 0 45 0},clip]{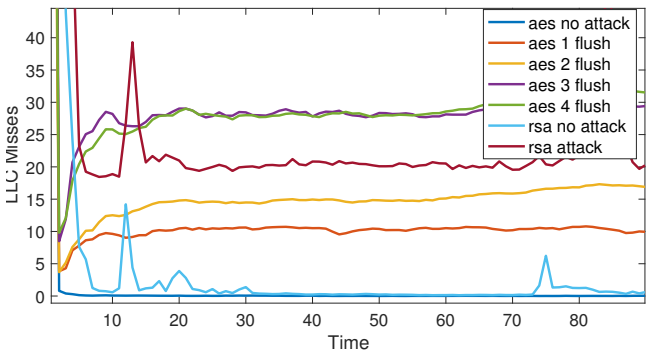}
\vspace{-8mm}
\caption{Mean LLC miss traces over time for AES and RSA executions in the presence and absence of cache attacks. The numbers next to \emph{flush} indicate the number of lines flushed at a time. After the start up peaks, the misses go to zero in the absence of a cache attack, while under attack they remain high.}
\label{time_attacks}
\vspace{-4mm}
\end{figure}

During our initial experiments we have observed that all implementations feature a \emph{start up} behavior, where data is loaded into the cache for the first time and the frequency of the CPU might be adjusted. The subsequent executions feature a more constant behavior. 
Regarding to the counters analyzed, for AES these start up executions show indistinguishable behavior under attack and without an attack. For RSA, they can be distinguished, but it would be necessary to know exactly if the current sample belongs to the start-up group of \emph{normal} executions. 
However, if we switch to continuous monitoring, the differences between algorithms disappear and the start-up behavior is restricted to a short time at the beginning of the processes. Figure~\ref{time_attacks} represents the mean value of the L3 miss counter in our initial scenario setup, for \texttt{Flush+Reload} attacks as well as \emph{normal} execution of the mentioned encryption processes. The average is computed over 1000 encryptions and counters are read every 100 $\mu s$. It can be clearly observed that after the initial transient state, the number of misses goes to zero in the absense of attacks (\texttt{aes no attack} and \texttt{rsa no attack}) for both crypto primitives. It can also be observed that the mean number of misses in the case of an attack varies with the number of lines flushed each time \texttt{aes 1 flush, aes 2 flush...}. Thus, with concurrent monitoring, both algorithms behave similarly for the \emph{normal} executions.

Switching to continuous monitoring of the counters implies that the information on total encryption times or reference cycles is no longer useful nor available. 
To ensure the information of the other counters mentioned in Table~\ref{tab:HPCEval} is still optimal for attack detection, 
we performed a new analysis considering each sample collected at a period of 1\,ms as an independent input to the selection attribute algorithms.  The results show that for the LLC misses counter the \texttt{infoGain} increases up to 0.92, while values for the other counters decreases. Additionally, the relief algorithm output still gives better score for the L3 cache misses (0.18) and in this scenario, this value is still 5 times bigger than the weight of the next counter, indicating the \texttt{L3\_TCM} is still the one counter of choice for cache attack detection.

We performed additional experiments to determine how well a cluster algorithm would distinguish between attacks and non-attacks with the periodically sampled data from several counters at once. WEKA also includes clustering algorithms. We tested \emph{EM} and \emph{Self Organizing Maps}, setting the number of clusters to two. The most interesting result of this experiments is that while these algorithms were able to classify in the same cluster respectively 84\% and 91\% of the \emph{attack} samples when using only the LLC misses counter, this number decreases to around 50-60\% when adding other counters. 
These results indicate that cache attacks can be detected, regardless of the algorithm the victim process runs, by only using information gathered from the L3 cache miss counter. The algorithms feature zero misses after the initial warm-up, except if an attacker is forcing misses. Additionally, as all known cache attacks, including \texttt{Flush+Flush}, cause cache misses on the victim process to obtain information, the results obtained here for the \texttt{Flush+Reload} attack are applicable for other attacks. Thus, we decided to only use this one attribute, as it provides most information and, also allows us to keep the detection tool simple.

\section{Cache Shield}\label{sec:goals}

So far, techniques proposed to detect cache attacks imply monitoring the victim VM, the attacker VM, and any other VM running in the same host~\cite{ChiappettaEtAl2016,Zhang2016}. Monitoring all VMs at rates which vary from 1\,us to 5\,ms result in huge overheads, and increases with each new virtual machine allocated in the same host. 

As a consequence, cloud providers may not want to implement such a tool, as it increases overall system cost, while the benefit of preventing cache attacks might be a benefit only few customers are willing to pay for. Yet, only the hypervisor, and thus the cloud service provider (CSP) has the ability to monitor \emph{all} VMs on a system. Indeed, as of now, we are not aware of any CSPs employing VM monitoring for microarchitectural attacks. 
 
As a difference with previous approaches, our goal is to design \CS\ in such a way that we avoid monitoring all the other processes or VMs running in the same host, i.e., we only focus on our own process. We assume that we have access to the performance counters within the VMs. Although most cloud providers currently do not allow access to the performance counters, hypervisor systems such as VMware and KVM can be easily configured to permit reading the counters inside the VM. Moreover, it is possible to decide which of the VMs allocated in a host would have access to the counters for their processes upon request. Even when our approach can be implemented at the hypervisor, we believe that for cloud providers would be easier just to enable the counters for the VMs that require it, leaving the responsibility on them, than to take care of these attacks.

By leaving the choice of deciding which processes should be monitored and when in the hands of the user, the impact in performance of such monitoring is reduced to a minimum, as we only watch a possible victim when it is executing the protected task. From the cloud provider's point of view, this way of facing detection also means no waste, as it only affects the implied VM and only when it is necessary. Additionally, as the user decides when it is necessary to protect a process, we avoid the need to detect when a sensitive process is executed. As a consequence, we also reduce the risk of not detecting the execution of this sensitive process and then the probability of missing an attack.

\begin{figure}[tb]
\centering
\includegraphics[width=0.4\textwidth]{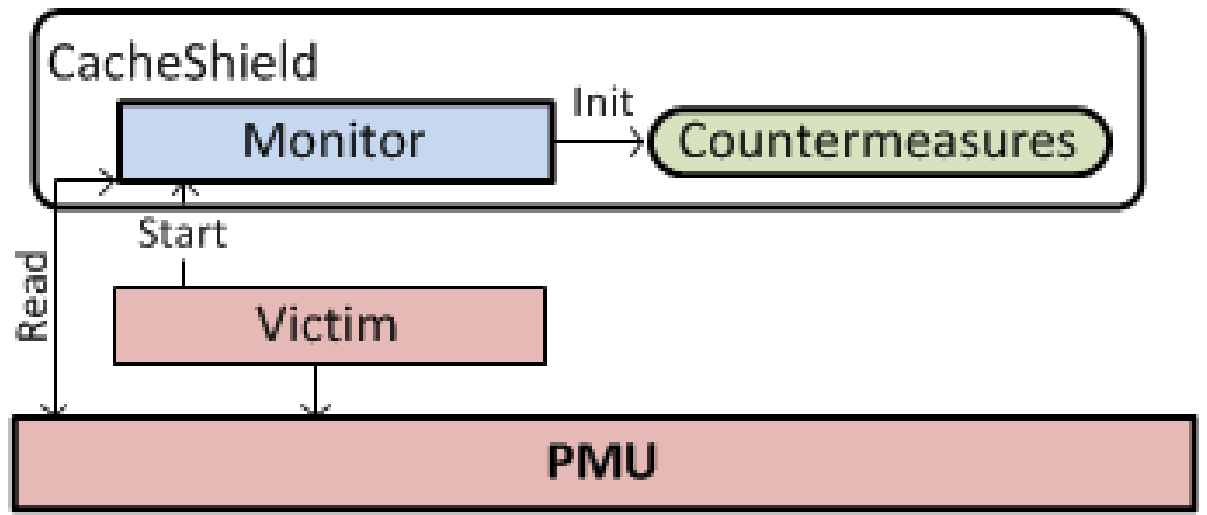}
\vspace{-3mm}
\caption{Overview of \CS .}
\label{fig:overview}
\vspace{-6mm}
\end{figure}

Figure~\ref{fig:overview} presents a diagram of our proposed solution. Whenever a user wants protection, he informs the \CS\ module, which utilizes the information gathered from the performance counters to decide whether the user is being attacked. If \CS\ detects an attack, an appropriate response mechanism to prevent the information leakage is put in place. Although we mainly focus on the detection phase, we discuss in Section~\ref{sec:counter} some of the countermeasures that can be implemented to effectively prevent the attack from retrieving information, such as the utilization of a fake key or the addition of noise patterns in the cache.

\subsection{Detection Algorithm}

One of our goals is obtaining a technique for attack detection no matter which algorithm is being attacked. Additionally, we want to detect all types of cache attacks, even unknown attacks, for which the tool has not been trained to detect. 

Supervised learning algorithms such as neural networks, have already been used to detect certain cache attacks. As any supervised algorithm they have to be ``trained'' to detect the attack. That is, they require a labeled data set including data from the different attacks we want to detect, so they can build models of them and identify their characteristic features. 
The drawback of supervised learning is precisely that we need to train the algorithm for each situation, for each algorithm and each attack. As a consequence new attacks, or attacks with different patterns would not be detected.

The alternative is using unsupervised techniques. An unsupervised algorithm does not receive labeled data, by itself tries to cluster the received data into different groups or to find relationships between different inputs in order to put any new sample in the appropriate cluster. We will briefly explore clustering techniques in the next section to select the counters which can identify an attack. Other kind of unsupervised techniques are anomaly-based detection algorithms, which in theory could detect ``zero-day" attacks. To the best of our knowledge no successful cache detection fully based on anomaly-detection techniques has been yet demonstrated.

Change-point detection methods are designed to deal with the problem of detecting abrupt changes in distributions. Under the assumption that cache attacks have an effect in the performance of the protected algorithms, change-point detection algorithms stand as great candidates to detect LLC attacks. We propose an algorithm based on change point-detection techniques which is self-learning so it adapts itself to detect different attack patterns, which allows us to fix the attack detection delay, and which is computationally simple so it respects the constraint of minimum impact in performance and can be implemented online. 

\subsection{Cache Shield Design}
\CS\ monitors the counters for LLC misses and for total cycles. The former gives information about the use of the LLC of the protected processes while the latter gives information about when it is running or when it has finished. Based on Figure~\ref{fig:overview}, the \CS\ module needs the PID of the process we desire to protect and the process protected also needs to know the PID of the \CS\ process. The reason is both processes need to communicate with each other (one needs to inform the other when to watch and other needs to inform the one when there is an attack going on), and that the counters can be attached to gather the data from a single process given its PID. 

On Unix systems, the easiest way to use \CS\ is to use the fork operation, and then to use the \textit{exec} system call to run the module and to give it the PID of the parent process. The parent process then can execute the desired operation while being monitored. In case that the parent process stops or waits for something, \CS\ automatically stops after noticing the parent has not been running for a while. This means that when the parent runs again, it needs to send a ``SIGCONT'' signal to the \CS\ tool. In a similar way, if the tool detects an attack, it can send a signal to inform the parent. On Windows Systems the mechanisms for inter-process communications are slightly different, but the tool can be also adapted.

\medskip\noindent\textbf{Change Point Detection:} 
In order to effectively asses the detection task, we made use of change point detection theory (CPD)~\cite{Basseville:1993:DAC:151741}. This theory can be used to construct the commonly known as quick detection algorithms, which have been successfully applied for quality control, signal-processing, anomaly or intrusion detection tasks among other problems~\cite{1677904,NAV:NAV3220370504,Tartakovsky2006252,MontesDeOca20101288}. The assumption in these scenarios is that the parameters describing the monitored system do not change or change very slowly under normal conditions. The parameters can, however, change at unknown time instants (including at startup) into anomalous conditions. 
Thus, CPD algorithms are used to determine if there has been a significant change in the characteristic parameters of the monitored system, quickly and with high confidence.

The theory of change point detection leads to the development of efficient algorithms presenting certain optimality properties, in the sense that for a given false-alarm rate (FAR) they minimize the average time it takes to detect the change in the descriptive system's features~\cite{MontesDeOca20101288}. CPD algorithms can be easily implemented, do not require too much memory and, as a consequence do not have significant computation overhead. These methods belong to the ``anomaly detection'' class and are \emph{unsupervised} techniques. Hence, they are well-suited to detect new attacks. All these properties made them very attractive for the attack-detection objective. 
In the following, we describe the parameters of the algorithm and how we assess key issues, such as the choice of models or the use of prior information.

We denote the sequence of observations of the $N$ variables monitored in parallel as $X(t)=(X_{1}(t),...,X_{N}(t)), t \geq 1$. Before a change occurs, the joint probability distribution (pdf) of the random variables $X_{1},...,X_{N}$ also known as \emph{prechange distribution}, can be denoted as $p_0(X_{1},...,X_{N})$. If a change occurs at an unknown time instant $\lambda$, the observations will follow a different distribution $p_1(X_{1},...,X_{N})$, also called \emph{postchange distribution}. That is, when $t < \lambda$ the observations $X(t)$ will have conditional pdf $p_0(X(t)|X(1),...,X(t-1))$, and pdf $p_1(X(t)|X(1),...,X(t-1))$ for $t \geq \lambda$.  

Under the hypothesis that a change has occurred, the stopping time $\tau$ at which the alarm is triggered gives a measurement of the detection time. It is typically defined as the first time the change sensitive statistic watching the system, exceeds a threshold. Naming $E_0$ and $E_\lambda$ the expectations for the sequence of observations prior and after the change at time $\lambda$ , the average detection delay (ADD) is defined as:

\begin{center}
$ADD_\lambda ( \tau ) = E_\lambda ( \tau - \lambda | \tau	\geq \lambda)$
\end{center}
On the other hand, considering that there has not occurred any alarm, the mean time between false alarms will be given by the expression $E_0 \tau $. As a consequence of this definition, the average frequency of false alarms or false alarm rate (FAR) is defined as:
\[
FAR(\tau)=\frac{1}{E_0 \tau }
\]

For a good detection procedure it is expected low FAR and small values of the expected detection delay. The design of CPD algorithms often involves a trade-off between these two parameters. Page's cumulative sum (CUSUM) detection algorithm~\cite{page1954continuous} is one of the most popular CPD algorithms: with a full-knowledge of the pre-change and post-change distributions it provides an optimal scheme minimizing the worst-case detection delay. Page's CUSUM algorithm utilizes the log-likelihood ratio (LLR) to check the hypothesis that a change occurred, LLR is defined as:  

\begin{center}
$s(t)= ln \dfrac{p_1(X(t)|X(1),...,X(t-1))}{p_0(X(t)|X(1),...,X(t-1))}$
\end{center}

The key property of this ratio is that a change in the parameter under study will also cause a change in the sign of the log-likelihood ratio. In other words, $s(t)$ shows a negative drift before change and a positive drift after change. The relevant information for the detection task lies then in the difference between the value of $s$ and a minimum value. The decision rule is based on a comparison with a threshold $h$: 

\begin{center}
$g_k= S_k - m_k \geq h$
\end{center}
where  
\begin{align*}
S_k &=\sum_{t=1}^{k}s(t) & m_k &=\min_{1\leq j \leq k} S_j 
\end{align*}

This decision rule can be replaced by the following, which obeys the recursion and whose value for the initial observation is $k=0$.

\begin{center}
$g_k = \max \left \lbrace 0,g_{k-1}+\ln \dfrac{p_1(X(k))}{p_0(X(k))}\right\rbrace \geq h$
\end{center}
Then the detection time for the given threshold is 
\begin{center}
$\tau(h) = \min \left \lbrace k \geq 1 : g_k \geq h \right\rbrace $
\end{center}

Although this first approach considers that both distributions are known, this assumption is usually not true, and as a consequence this proposal has to be adapted for each situation. We may know one distribution in advance or none, so it may be necessary to estimate the parameters of the algorithm during the runtime. As long as the estimators for the distributions and the real observation meet certain convergence conditions, we will be able to fix for example the desired detection delay or the FAR.

\medskip\noindent\textbf{ 
Change Point Detection in \CS :} 
While facing the cache attack detection, the attack may start from the very beginning or it may start after a few ``normal'' transactions. Both situations are efficiently managed by the proposed CUSUM algorithm. We assume that each new sample can be classified into one of two different groups or clusters, namely ``attack'' and ``non-attack''. The ``non-attack'' cluster represents how we expect the protected process to behave under normal conditions.
Based on the information we can gain from the counters, this assumption is that after a few samples corresponding to the initialization of the protected process, the number of L3 cache misses will be around 0, then $\mu_{na}=0$. On the other hand, when there is an attack, we have observed that the mean number of misses is $\mu_a$. Then each new sample belonging to the ``attack'' cluster will be around $\mu_a$. The value of $\mu_a$ is unknown and depends on the attack so it needs to be computed and recalculated with each new sample.

If we denote as $miss_i$ each new sample that the \CS\ module gets referring to the protected process, we need to decide if it belongs to one cluster or to the other. To do so, we compute the value of the "probability" that $miss_i$ belongs to each one making use of the distance metric, this way we define the distance from $miss_i$ to $\mu_{na}$ as: 

\begin{center}
$d_{na}(i) = miss_i -\mu_{na}=miss_i$ 
\end{center}

Then, the distance with the "attack" cluster will be 

\begin{center}
$d_{a}(i) = | miss_i -\mu_{a} |$ 
\end{center}

As stated before the value of $\mu_a$ is unknown when we start to monitor the process. We select an arbitrary initial value, and whenever a new sample $miss_i$ is obtained, if $miss_i  \geq 0$ we update the value of $\mu_{a}$ as follows:

\begin{center}
$\mu_{a} = (1-\beta) * \mu_{a} + \beta * miss_i $ 
\end{center}

This method is known as exponentially weighted moving average, where the weight of the older datum decreases exponentially. This way of estimating the mean of the "attack cluster" makes the the election of the initial arbitrary value irrelevant after collecting a few misses samples. If the initial value is chosen too low, we may trigger false positives. 
We recommend the election of an initial value higher than 10, in order to keep the rate of false positives low, while being able to detect the attack in a reasonable time. We will further discuss the noise tolerance of the proposed detection algorithm in the next section. In our experiments we set $ \beta = 0.05$ and the initial value to 12.5.

Now we are in conditions to define the probability of belonging to each cluster:
\begin{align*}
   p_{na}(miss_i ) &= \dfrac{d_{a}(i)+1}{| d_{na}(i) |+| d_{a}(i) |}, & 
 & p_{a}(miss_i ) &= \dfrac{d_{na}(i)+1}{| d_{na}(i) |+| d_{a}(i) |} 
\end{align*}
The value 1 has been added to avoid divisions by 0 in the LLR calculation that has to be performed as part of the detection algorithm. As a result, for every sample $k$, $k \leq 1$ we can express the detection rule as follows:
\begin{center}
$g_k = \max \left \lbrace 0,g_{k-1}+\log \dfrac{d_{na}(k)+1}{d_{a}(k)+1}\right\rbrace \geq h$
\end{center}

As it can be easily derived from the previous equation and according to the properties of the LLR, when the number of misses is 0 or close to 0, the distance between the sample and the "non-attack" cluster $d_{na}(k)$ will be lower than the distance to the attack cluster $d_{a}(k)$, so the value of the metric $g_k$ decreases or stays at zero. On the other hand, readings from the LLC misses counter approaching to the attack cluster will increase the value $g_k$. The properties of this approach let us choose the threshold based on a minimum detection time we want to achieve. Note that when the error in the estimation of the mean $\epsilon$ approaches to zero, $d_{a}(i)=\epsilon$ also tends to zero, then the increase in the value of $g_k$ is also limited
\begin{center}
$\log \dfrac{d_{na}(k)+1}{d_{a}(k)+1} \leq \log ( \mu_{a} +1) $
\end{center}

As a consequence, the minimum expected detection time for the given threshold $h$ is:
\begin{center}
$\tau_{e}(h) \geq  \dfrac{h}{\log ( \mu_{a} +1) }$
\end{center}

or reformulating this equation, the threshold h, for a minimum expected delay $\tau_{e}$

\begin{center}
$h_{\tau} \leq  \tau_{e}*\log ( \mu_{a} +1) $
\end{center}

The unit of the $\tau_{e}(h)$ is \emph{number of samples}. 
Given that the most effective cache attacks can potentially extract most of the key with just one execution of the victim, the sampling rate must be chosen lower than the execution time of the victim. 
As the execution time of these algorithms is in the order of few milliseconds, a sampling rate of 100 $\mu s$ seems sufficient to provide evidence of the attack. This frequency can be increased at additional load for the system. So, for an expected detection delay of 1\,ms with a sampling rate of 100 $\mu s$ we can define the threshold as $h=10*\log ( \mu_{a} +1) $. As a result of this selection of $h$, when the $\mu_{a}$ is recalculated, the threshold should be recalculated too. The choice of the threshold $h$ also determines the tolerance to noisy frames, and as a consequence the false positive rate. In practice, the false positive rate cannot be estimated and has to be measured. 

\begin{figure}[tb]
\centering
\includegraphics[width=0.49\textwidth,trim={55 0 45 0},clip]{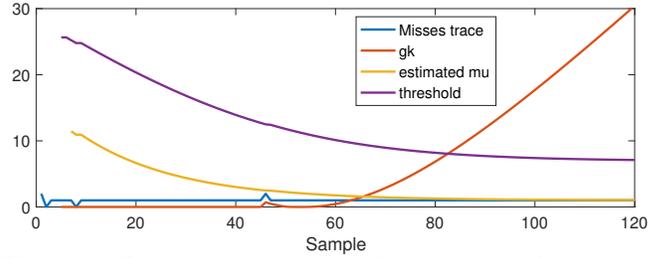}
\vspace{-8mm}
\caption{Relevant parameters for the detection task, \texttt{prime+probe} attack on AES}
\label{fig:detection_plot}
\vspace{-6mm}
\end{figure}

Algorithm \ref{cashield} summarizes CacheShield implementation and an example of the values of the parameters considered in the detection process is given in Figure~\ref{fig:detection_plot}.

\renewcommand{\algorithmicrequire}{\textbf{Input:}}
\renewcommand{\algorithmicensure}{\textbf{Output:}}
\algnewcommand{\algorithmicendif}{\textbf{end if}}
\algblockdefx[IF]{If}{EndIf}[1]{\algorithmicif\ #1\ \algorithmicthen}{\algorithmicendif}
\algnewcommand{\algorithmicendfor}{\textbf{end for}}
\algblockdefx[FOR]{ForAll}{EndFor}[1]{\algorithmicfor\ #1\ \algorithmicdo}{\algorithmicendfor}
\begin{algorithm}
\caption{CacheShield detection algorithm}
\label{cashield}
\begin{algorithmic}[0]
\Require \textbf{Process PID}
\Ensure \textit{\textbf{Attack detected}}
\State \emph{read\_counters}(misses,cpu\_cycles);
\State \emph{wait;}
\While{\emph{victim\_is\_running}}
	 \State \emph{read\_counters}(misses,cpu\_cycles);
     \If {misses$>0$}
     		\State \emph{update} $\mu_{a}$;
     		\State \emph{update} $h$;
     \EndIf
     \State \emph{calculate} $g_k $
     \If {$g_k > h$}
     		\State \emph{trigger\_alarm};
     \EndIf
     \State \emph{wait;}
\EndWhile
\State \textbf{return} \textbf{detected};
\end{algorithmic}
\end{algorithm}

\section{Evaluation of \CS}\label{sec:eval}

Once we have defined the relevant parameters of the detection algorithm and described it in detail, we evaluate its performance. To this end we ran several experiments in different environments and machines. 
\begin{description}[leftmargin=4ex]
\item[Native Environment] The experiments for non-virtualized environments were performed in an Intel Core i7-4790 CPU 3.60GHz machine with 8\,MB of L3 cache and 8\,GB of RAM, with Centos 7 OS.
\item[KVM-based Hypervisor] These experiments used the same hardware as above, but this time within a VM also with Centos 7 hosted in KVM as hypervisor.
\item[VMware-based Cloud Server]  We have also executed experiments in a host managed with VMware, this machine is equipped with a Intel XeonE5-2670 v2 processor, 25Mb of L3 cache and 32GB of RAM. The OS in these VMs was Ubuntu 12.04.
\end{description}

\begin{table*}[bth]
\caption{Mean detection time (ms) per attack and scenario for the evaluated crypto algorithms. Note that in all cases \CS\ has the same configuration and that detection times are much lower than the ones required for the attack to succeed }\label{tab:times1}
\vspace{-3mm}
\centering

\begin{tabular}{lcccccccccc}
\toprule
\textbf{Scenario}    & \multicolumn{4}{c}{\textbf{AES}} & \multicolumn{3}{c}{\textbf{RSA}}  & \multicolumn{3}{c}{\textbf{ElGamal}}\\ 

 &\textbf{ F+R (1)}& \textbf{F+R (4)} & \textbf{F+F (1)} &\textbf{ P+P}~~ & ~~\textbf{F+R} & \textbf{F+F} & \textbf{P+P}~~ & ~~\textbf{F+R} & \textbf{F+F} & \textbf{P+P} \\
\midrule
Native  &  3.98 & 4.48 & 3.38 & 5.08 & 3.70 & 3.54 & 5.16 & 2.97 & 3.47 & 3.68 \\

KVM & 7.38  & 7.05 & 6.64 & 9.53 & 4.08 & 3.92 & 4.93 & 3.76 & 3.45 & 3.98 \\

Vmware &8.75 & 5.98 & 10.74 & 13.42 & 4.43 & 3.87 & 4.51 & 4.83 & 5.06 & 7.08\\

  \bottomrule
\end{tabular}
\vspace{-3mm}     
\end{table*}

When a user is executing the crypto algorithm in their own machine, they can get information about the utilization of such machine or other task running concurrently. However, when executing the crypto algorithms in cloud environments they can not get any information about what their neighbors are doing. In such scenarios, it becomes mandatory to study how the execution of different applications running in parallel with the protected process affects the behavior of \CS. Note that as we use the "total cycles'' counter (to determine if the victim is executing or not) and the LLC misses counter to decide if there is an attack going on, applications consuming high amount of memory resources are the most likely to cause the LLC misses indicator to rise, and as a consequence, to trigger false positives. For this reason, we have selected several worst case scenario applications with high memory activity to run in parallel with the \emph{victim} and \CS:
\begin{description}[leftmargin=4ex]
\item[Yahoo Cloud Serving Benchmark] This benchmark was originally designed as a tool that provides a common evaluation framework and a set of common workloads to test the performance of different serving stores as elastic search, Cassandra, MongoDB among others~\cite{Yahoo}. It allows different configurations for the workloads and provides a set of \emph{example} workload scenarios, together with a workload generator, which generates the load to test storage systems. In our experiments, we use this benchmark with the Apache Cassandra database and the example workload named \emph{workloada}.
\item[Video Streaming] Another kind of application that can generate cache misses is web-browsing or video streaming. The video streaming VM continuously streams and plays back youtube videos on the firefox browser.
\item[Randmem Benchmark] This benchmark was originally intended to test the impact of burst reading and writings~\cite{randmem}. Depending on the configuration, the benchmark accesses data stored in an array either sequentially or in random order. The tool also allows to configure the size of the memory it is going to use, by default it tries to use as much as possible, up to 2\,Gb. In our experiments, we launch each \texttt{randmem} instance with no memory limitation, which means 2\,Gb of RAM memory are used by each instance.
\end{description}

To show the applicability of \CS\ to a broad range of implementations that require protection against cache attacks, we chose from a range of crypto primitives and implementations, though focusing on vulnerable ones, since such legacy implementations actually require protection. The three crypto algorithms considered as \emph{victims} are
\begin{description} [leftmargin=4ex]
\item[AES] as the most common symmetric encryption algorithm. We consider the T-Table implementation of AES from Openssl 1.0.1f, which is fast, but also leaky.
\item[RSA] is the probably most widely used signature and public key encryption algorithm. We analyzed the RSA implementation from Openssl 1.0.1f, with a  2048 bit key, and the RSA\_\-FLAG\_\-NO\_\-CONST\-TIME flag set. 
\item[ElGamal] we chose the ElGamal implementation of libgcrypt 1.5.0 with a 4096 bits key. Unlike AES and RSA, ElGamal was not considered during the design of \CS, and hence shows how \CS\ can be expected to perform for other types of algorithms.
\end{description}
These algorithms differ quite significantly in their particular implementation and usage of cache. Many other potentially leaky codes might require protection, and we are confident that \CS\ will perform well.

\begin{table}[tb]

    \caption{False positive rate for different scenarios and algorithms. (Instances: \textbf{Y} - Yahoo Cloud Serving, \textbf{V} - Video Streaming; \textbf{R} - Randmem)}
		\label{tab:noise_all}\vspace{-3mm}
		
\resizebox{0.98\columnwidth}{!}{%
\centering
\begin{tabular}{lcccrrr}
\toprule
\textbf{Scenario} & \multicolumn{3}{c}{\textbf{Noise Instances}} & \multicolumn{3}{c}{\textbf{False Positives}}\\ 
 & \textbf{~~Y~~}  & \textbf{~~~V~~~}  & \textbf{~~R~~} & \textbf{AES} & \textbf{RSA} & \textbf{ElGamal}\\ 
\midrule
KVM & 1 & 0 & 0  & 1.2\%  & 4.5\% & 2.8\% \\ 
KVM & 0 & 1 & 0 & 1.1\%  & 3.4\% & 0.6\% \\ 
KVM & 0 & 0 & 1 & 12.2\%  &  21.4\% & 15.4\% \\ 
Native & 1 & 0 & 0 & 1.2\%  & 4.1\%  & 2.4\% \\ 
Native & 0 & 1 & 0 & 0.5\%  & 1.3\% & 0.3\% \\ 
Native & 0 & 0 & 1 & 11.1\%  &  19.2\% & 13.8\% \\ 
VMware & 1 & 2 & 10 & 0.1\% & 5.9\% & 4.1\% \\ 
\bottomrule
\end{tabular}}
\vspace{-3mm}
\end{table}

To evaluate the effectiveness of \CS\ across different types of cache attacks, we implemented and performed three popular attacks, namely \texttt{Flush+Reload}, \texttt{Flush+Flush} and \texttt{Prime+Probe}. We collected data for the above-mentioned algorithms under attack as well as from normal executions, as baseline behavior. 
Under each configuration, we collect data for more than 1000 executions of the crypto primitives, and in the case of the AES attack we also consider different attack rates (number of lines flushed at a time), as the attacker may try to gain different amount of information from the T-tables per execution~\cite{DBLP:journals/corr/GreenLZIHE17}. As stated in previous sections, the main characteristics defining the detection algorithm are the mean detection time, and the false positive rate. Table~\ref{tab:times1} presents the results for mean detection time under different configurations, for the different attacks and algorithms and table \ref{tab:noise_all} shows the results related with false positives in noisy environments.

\begin{figure}[tb]
\centering
\includegraphics[width=0.49\textwidth,trim={60 0 45 0},clip]{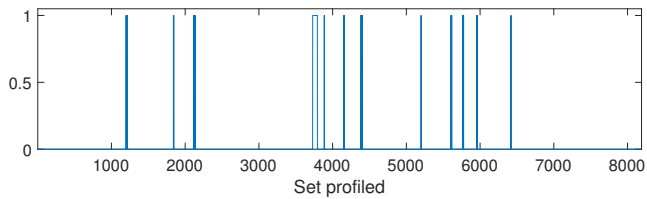}
\vspace{-6mm}
\caption{Output of \CS\ when the cache is profiled accessing each set. "1" indicates a positive attack detection.}
\label{cache_profile_plot}
\vspace{-5mm}
\end{figure}

Note that the attack requirements for \texttt{Flush+Flush}/\texttt{Flush+Reload} and \texttt{Prime+Probe} differ significantly. While \texttt{Flush+X} attacks are faster and more precise, they require shared data, i.e. deduplication between attacker and victim. All the attacks performed in virtualized scenarios were across VMs so we enabled deduplication features (KSM and TPS for KVM and vmware respectively) to perform \texttt{Flush+X} attacks. \texttt{Prime+Probe} attacks work across VMs even without deduplication, so we disabled deduplication and enabled huge pages. \texttt{Prime+Probe} attacks require, prior to the information extraction, a \emph{profiling} of the cache~\cite{lastlevel,sca,Inci2016}. The profiling stage reveals the sets the victim process is accessing and that carry the necessary information to succeed in the attack. In this situation, the detection tool will trigger an alarm whenever the set being tested by the attacker was actually used by the victim. Fig.~\ref{cache_profile_plot} visualizes the output of the detection algorithm,for the cache profiling stage of an 8\,MB L3 cache when the target is the T-table implementation of AES. The x-axis represents each set of the cache being evicted during the \texttt{Prime+Probe} profiling step; a 1 on the the y-axis indicates that an alarm has been triggered. Thus, alarms are only triggered when the cache attack affects the target. 

For all evaluated attacks, the detection rates are 100\%. Note that the sampling rate is 100 $\mu$s and that we want to detect the attack before the end of each decryption (for public key cryptography). 
If we wished to detect attacks against algorithms whose duration is below 5 or 6 ms, we will need to increase the sampling rate, since mean number of samples required to detect the attack cannot be lowered arbitrarily without increasing the FAR too much.
The duration of the decryption/encryption depends on frequency of the processor, and as a consequence on the machine. For example, ElGamal encryption takes around 11 ms when being attacked on the i7 machine, while this time increases up to 24 ms on the Xeon machine. Thus, we are able to detect attacks against ElGamal when less than the 37\% of the encryption has been performed for the i7 machine, and 30\% for the second machine in the worst case. Regarding to RSA this mean execution times are around 18 ms for the i7 machine and around 37 ms for the other. Then, in the worst case, on average we detect the attacks with less than 50\% of the decryption performed in the first case and with about 37\% of decryption in the second one.

\begin{figure}[t]
\centering
\includegraphics[width=0.49\textwidth,trim={40 0 45 0},clip]{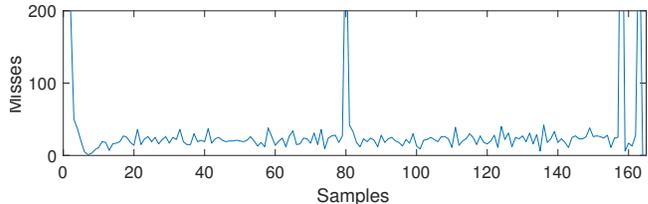}
\vspace{-8mm}
\caption{LLC misses for a noisy RSA execution under randmem benckmark.}
\label{rsa_noisy_miss}
\vspace{-4mm}
\end{figure}

\begin{figure}[t]
\centering
\includegraphics[width=0.49\textwidth,trim={45 0 45 0},clip]{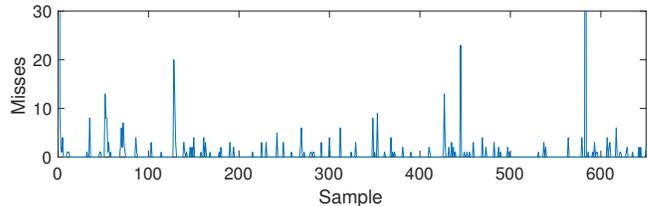}
\vspace{-8mm}
\caption{Sample of a noisy execution of AES under randmem benchmark.}
\label{noise_plot}
\vspace{-6mm}
\end{figure}

Regarding to the existing differences between false positive rates for AES and public crypto algorithms, these are easy to explain. While between AES encryptions exists some time in which the processor does nothing, the others execute uninterruptedly. This fact increases the probability of other processes accessing the cache during the same interval. For example, while the AES encryption in the period of 100 $\mu$s is only active during around 7000 cycles while the RSA process is active during about 30000 cycles for the VMware machine when there is no attack.  Fig.~\ref{rsa_noisy_miss} depicts the LLC misses for one noisy RSA encryption, besides the initialization steps, it can be observed a high amount of cache misses during the whole decryption. Similarly, Fig.~\ref{noise_plot} corresponds to one process performing AES encryptions.


The results also show that the tolerance to noise of the detection algorithm is more dependent on the hardware than on the virtualizing technology: While the results for the native and KVM scenarios are similar and the hardware is the same, the results are significantly better on the Xeon machine. The Xeon machine did not trigger any false positives when there were one or two VMs generating "noise" concurrently, until we launched several more instances. As this machine is more similar to the kind of machine cloud providers utilize, these results show that the tool is practical in these environments. 

One approach to reduce the false positives in noisy environments could be considering the variance of the samples collected in the CUSUM algorithms proposed, as attacks present low variance compared with noise. However, we could fail to detect attacks masquerading as memory activity by generating different number of misses each time. 
Another consideration relative to memory utilization, and the false positives that are triggered when is high memory utilization is that \texttt{Prime+Probe} attacks need low memory activity to accurately locate the sets and to perform the attack, other way it renders much more difficult. On the other hand, \texttt{Flush+Flush} attacks are more tolerant to noise, but memory activity degrade its performance. So it is not likely that the attacker performs the attack in a situation where the memory is highly utilized. Additionally, the level of utilization of public clouds is low~\cite{6118751_Utilization}, so the assumption of high memory utilization in the considered cloud scenarios may not be realistic. As for using the tool in our controlled physical machine, we can get to know which is the level of utilization of the memory and then decide if it is worth it to change the parameters of the detection algorithm.

\begin{figure}[t]
\centering
\includegraphics[width=0.49\textwidth,trim={40 0 45 0},clip]{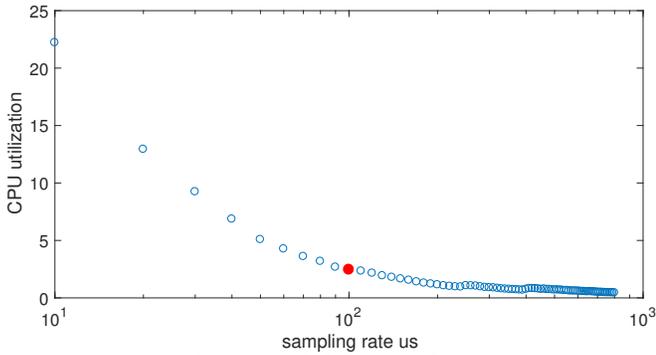}
\vspace{-8mm}
\caption{CPU utilization of the i7 machine in native environment for different sampling rates in microseconds. Highlighted the 100 us rate as it is the one we use in our experiments}
\label{i7_cpu}
\vspace{-4mm}
\end{figure}

\begin{figure}[t]
\centering
\includegraphics[width=0.49\textwidth,trim={40 0 45 0},clip]{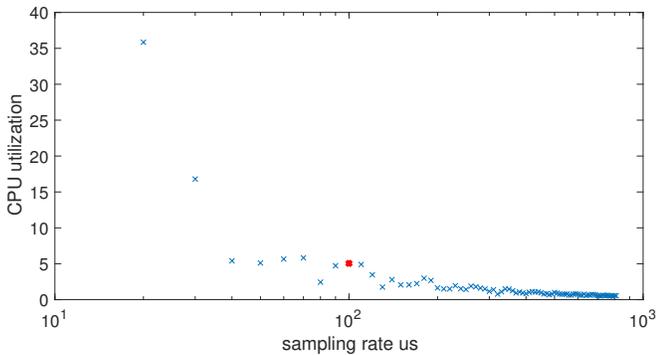}
\vspace{-8mm}
\caption{CPU utilization of the Xeon machine in virtualized environment for different sampling rates in microseconds. Highlighted the 100 us rate as it is the one we use in our experiments}
\label{vmware_cpu}
\vspace{-6mm}
\end{figure}

One last consideration about our tool is the amount of CPU it utilizes to monitor the victim and compute the detection algorithm. Fig. \ref{i7_cpu} and \ref{vmware_cpu} show the mean CPU utilization of \CS for different sampling rates and for different situations, namely when the victim is attacked and when is not, because the amount of operations it has to do changes, and again for both architectures, depending on the sampling rate. To obtain the CPU utilization we have measured the time it takes to read the counters and perform the calculations and the total time elapsed, then the utilization is given by its division. Note that sampling rates of 10 $\mu$s are not always achievable as sometimes (around 10\% of the time) it takes more time to read the counters and perform the calculations. Note that in both cases total utilization of our tool is below 5\% of CPU utilization.


\section{Cache Attack Countermeasures}\label{sec:counter}

Once an attack has been detected, \CS\ needs to react in some way. One way is to simply interrupt the monitored process and to purge used keys. While this approach ensures high security, it decreases the usability, as any false positives will result a total cryptosystem shutdown. An alternative is to continue execution, but to apply preventative measures to reduce or prevent the exploitability of the cache. 

\begin{description}[leftmargin=4ex]
\item[Adding Noise] A simple method to hinder cache attacks is making the channel noisier, e.g. through frequently flushing cache lines used by the protected process, or by performing additional reads on data. This approach works particularly well if critical data is known, e.g. the tables of an AES implementation. 
\item[Dummy Operations] An alternative approach is to perform dummy operations on meaningless secrets. In practice this can mean to run the protected process with a newly generated secret. the original process can either be paused, or be continued in parallel to the dummy process. Parallel processing obfuscates the true leakage. However, depending on the attack type, an attacker might still succeed with an increased number of observations. Pausing has the advantage that the attacker might actually extract the dummy key and discontinue the attack. The monitor can then restart the original process in the absence of the attack. 
Either way, the performance degradation is not negligible, but it only is incurred in the presence of an attack. 
\item[Protected Implementations] The main reason why leakage is still observed in security solutions is the performance overhead that pure constant time implementations present. A way of avoiding such a scenario is to use protected implementations \emph{only} when \CS\ detects an attack behavior. When no attack is detected, faster (less secure) implementations can be used.
\end{description}
Other more sophisticated solutions are also possible, but might not be as universally applicable. Since our focus is on the lightweight detectability of cache attacks at the user level, we do not explore these additional avenues of countermeasures. \

\section{Conclusion} \label{sec:conclusion}

In this work we have introduced CacheShield, a tool that is able to detect all known types of cache attacks targeting cryptographic applications. The analysis of various hardware performance counters revealed that the LLC miss counter by itself carries enough information to detect cache attacks. We take advantage of change point detection algorithms and adapt them to our objective of cache attack detection. CacheShield was designed to detect attacks based on the characteristics of two particular algorithms, AES and RSA. The evaluation revealed that CacheShield can also be used for other algorithms (as shown for ElGamal) without further modification. It is also effective against ``unknown'' attacks, as all known attacks force cache misses on the victim. This behavior can be easily detected, since the number of L3 cache misses of crypto algorithms approaches zero after a brief initial warm-up. In addition, we have shown that CacheShield tolerates considerably high amount of noise only triggering a few false positives in machines similar to the ones cloud providers use.

Previously proposed cache attack detection tools work at the hypervisor level and also need to continuously monitor all untrusted and concurrently running processes or VMs, resulting in huge performance overheads and often have questionable detection rates for novel attacks such as \texttt{Flush+Flush}. \CS\ only needs access to the protected victim process, and only during its execution, greatly reducing the waste. 
All major hypervisor systems support transparent access to hardware performance counters for guest VMs while ensuring proper isolation between VMs. We urge Cloud Service Providers to enable these features in their systems and thus give their tenants finally the means to protect themselves against cache attacks with tools such as \CS .

\section{Acknowledgments}

Visit of Samira Briongos to Vernam group at Worcester Polytechnic Institute has been supported by a collaboration fellowship of the European Network of Excellence on High Performance and Embedded Architecture and Compilation (HiPEAC). This work was in part supported by the National Science Foundation under Grant No. CNS-1618837 and by the Spanish Ministry of Economy and Competitiveness under contracts TIN-2015-65277-R, AYA2015-65973-C3-3-R and RTC-2016-5434-8.

\bibliographystyle{ACM-Reference-Format}
\bibliography{ms} 

\clearpage

\end{document}